\documentclass[12pt,oneside,english]{amsart}
\usepackage[T1]{fontenc}
\usepackage[latin9]{inputenc}
\usepackage{geometry}
\usepackage{amstext}
\usepackage{amsthm}
\usepackage{graphicx}
\usepackage{setspace}
\usepackage{esint}
\makeatletter
\numberwithin{equation}{section}
\numberwithin{figure}{section}
\theoremstyle{plain}
\newtheorem{thm}{\protect\theoremname}
\theoremstyle{definition}
\newtheorem{defn}[thm]{\protect\definitionname}
\theoremstyle{remark}
\newtheorem{rem}[thm]{\protect\remarkname}

\usepackage{lscape}

\makeatother

\usepackage{babel}
\providecommand{\definitionname}{Definition}
\providecommand{\remarkname}{Remark}
\providecommand{\theoremname}{Theorem}

\begin{document}

\title{A Partition Theorem for a Randomly Selected Large Population}

\maketitle
\begin{center}

{\large{}Arni S.R. Srinivasa Rao}{\large\par}

Laboratory for Theory and Mathematical Modeling, 

Medical College of Georgia, 

Department of Mathematics, Augusta University, 

1120, 15th Street, AE 1015 

Augusta, GA, 30912, USA, 

Tel: +1-706-721-3786 (office). 

Email: arrao@augusta.edu 

{\large{}$ $}{\large\par}

\end{center}

\vspace{0.5cm}
\begin{abstract}
We state and prove a theorem on the partitioning of a randomly selected
large population into stationary and non-stationary components by
using a property of stationary population identity. Applications of
this theorem for practical purposes is summarized at the end. 
\end{abstract}

\vspace{0.5cm}



\section{Stationary Population}

Stationary population assumptions and related mathematical formulations
were of interest to Edmond Halley an astronomer to influential mathematician
Leonhard Euler to twentieth-century famous theoretical population
biologist Alfred Lotka. A population is said to be stationary if it
has a zero growth rate and a constant population age-structure, and
if not, the population is said to be non-stationary. Lotka associated
his theory of stationary populations with the average rate at which
a woman in her lifetime will be replaced by a girl (which we call
the net reproduction rate). In fact, Lotka \cite{Lotka1925,Lotka1939}
logically argued that the rate of natural growth, $r,$ of a population
will be zero in (\ref{eq: r=00003Dln(r0)}) when the net reproduction
rate of the population, $R_{0}$ will be equal to one (the notation
of $R_{0}$ was introduced by Lotka). The relation between $r$ and
$R_{0}$ is given by 

\begin{equation}
r=\frac{\ln(R_{0})}{L},\label{eq: r=00003Dln(r0)}
\end{equation}
where $L$ is the length of the generation of a population and is
expressed as,

\begin{equation}
L=\frac{1}{R_{0}}\intop_{0}^{\infty}xa(x)s(x)dx,\label{eq: length of generation}
\end{equation}
where $a(x)$ is the age-specific fertility rates of women of age
$x$ to give births to girl babies and $s(x)$ is the survival probability
function for the women to live up to the age $x$.

We define below stationary population identity (SPI) or the Life Table
Identity for a discrete population as given in \cite{RaoCareyOSPI}. 
\begin{defn}
Stationary Population identity (SPI) or the Life Table Identity: Let
$X$ be the set of elements representing proportion of population
at each age of a stationary population or a life table population
and $Y$ be the set of elements representing the remaining time units
to be lived at each age of this stationary population. We say SPI
holds if $X=Y.$ 

Let $A=\{0,a_{1},a_{2},...,\omega\}$ be the discrete set of ages
in the population of a life table. Here $\omega$ is the maximum age
in the population life table. The SPI holds for a life table means,
in a life table, the proportion of population at age $x$ (say, $f_{1}(x)$)
for $x\in A$ is equal to the proportion, (say, $f_{2}(x)$) of the
population who will live $x-$years, i.e. $f_{1}(x)=f_{2}(x)$$\:\forall\:x\in A.$ 
\end{defn}

This identity was theoretically demonstrated in partitioning large
populations \cite{RaoCareyOSPI}. This identity was frequently referred
also as life table identity in the mathematical population biology
and demography literature. Such identities are found in various published
literature, for example, see \cite{aMulleetal2004,bCarey2012,RaoCarey2015,Bouard,CareySilverRao}.
There were other related equalities in stationary populations which
suggest an average of a stationary population is equal to the average
expectation of remaining life (for continuous versions) \cite{Kim-Aaron (SIAM),CoxDR}.
Although Lotka and Cox in their respective works have chosen continuous
frameworks, in this article the population ages are treated on a discrete
framework. 

In this article, we have stated and proved a novel theorem that states
a criterion to partition a randomly selected population into stationary
and non-stationary components under a large discrete population model
framework. The criterion is to compare $f_{1}(x)$ and $f_{1}(x)$
and see if they are identical. Instead of the life table population,
the theorem suggests comparing the fraction population at each age
of the actual population at a time point with the fraction of the
populations that have remaining ages in the corresponding life table
population constructed at the same time. In general, we do not see
often stationary populations, except in life tables. But a careful
investigation at any large population data suggests that a sub-population
of a population could obey the properties of a stationary population.
Now, through the partition theorem that is proved in this article,
it becomes easier to understand the stationary component of large
populations instantly without being dependent on the measures like
NRR (net reproduction rate). Moreover, satisfying the theorem indicates
certainly one can find if any component of a large population is stationary. 

\section{Partition Theorem and Proof}
\begin{thm}
\emph{\label{prop:Stationary-Population-Identity}Stationary Population
Identity (SPI) partitions a randomly selected large population into
stationary and non-stationary components.}
\end{thm}

\begin{proof}
Let $P(t)$ be the randomly selected large population at time t and
$P_{x}(t)$ be its sub-population who are at age $x$, where $P(t)$
can be expressed as $P(t)=\int_{0}^{\omega}P_{x}(t)dx$ or $P(t)=\Sigma_{x=0}^{\omega}P_{x}(t)$
depending upon whether $x$ is measured as continuous or discrete
($\omega$ is the maximum age of life). In the current proof, we considered
P as a summation over discrete ages. Let $g_{1}(x)=\frac{P_{x}(t)}{P(t)}$
for all $x\in A$. We do not know whether $P(t)$ is a stationary
population or not. Stationary component of $P(t)$ (say, $M(t)$),
we define as, a sub-collection of various aged individuals of $P(t)$
who form a sub-population and satisfy $g_{1}(y)=f_{2}(y)$ for a set
of $y$ values in $A,$ where $f_{2}(y)$, as defined in the previous
section, the proportion of the population who will have $y-$years
remaining to live. Non-stationary component of $P(t)$ (say, $N(t)$),
we define as, a sub-collection of $P(t)$ who are of age $z$ and
satisfy $g_{1}(z)\neq f_{2}(z)$ for a set of all $z$ values in $A$.
The sum of the sizes of $M(t)$ and $N(t)$ will be $P(t).$

Let us choose all individuals at age $y$ in $P(t)$ and consider

\begin{equation}
g_{1}(y)=\frac{P_{y}(t)}{P(t)}.\label{eq:1}
\end{equation}

From the life table constructed for the population $P(t)$ (Assumably
for the single ages of the set $A$), we can obtain the proportion
of population who have expected remaining years within $[A,$ and
compare the proportion of population at each age $x\in A$ in $P(t).$
Note that, a life table is a mathematical model to synthetically demonstrate
the age-specific death rates of a population and to compute remaining
average years to live at each age in $A$. There are several books
available to understand details of life table constructions, for example,
see \cite{Wachter-book,Misra-BD,Preston-book}. That is, we will compare
$g_{1}(y)$ with another proportion $f_{2}(x)$ $\forall x\in A,$
where

\begin{equation}
f_{2}(x)=\frac{L_{x}(t)}{L(t)}.\label{2}
\end{equation}

Here $L_{x}(t)$ is the life table population at time $t$ whose remaining
years to live is $x$ and $L(t)$ is total life table population at
$t$. First, we match $\frac{P_{y_{1}}(t)}{P(t)}$ for a given $y_{1}\in A$
in (\ref{eq:1}) with $\frac{L_{x}(t)}{L(t)}$ for each $x\in A.$
If $\frac{P_{y_{1}}(t)}{P(t)}$ is equal to $\frac{L_{x}(t)}{L(t)}$
for some $x,$ then we call the corresponding life table proportion
of the population as $\frac{L_{y_{1}}(t)}{L(t)}$ $(say,$$f_{2}(y_{1})$).
That is, $\frac{L_{y_{1}}(t)}{L(t)}$ is the proportion of life table
sub-population who have $y_{1}$ years to live. This follows, 

\begin{equation}
g_{1}(y_{1})=\frac{P_{y_{1}}(t)}{P(t)}=\frac{L_{y_{1}}(t)}{L(t)}=f_{2}(y_{1}).\label{eq:3}
\end{equation}

Suppose $\frac{P_{y_{1}}(t)}{P(t)}$ does not equal to any of the
proportions $\frac{L_{x}(t)}{L(t)}$ for $x\in[0,\omega),$ then we
denote $y_{1}$ in (\ref{eq:3}) by $z_{1}$ and write this situation
as $g_{1}(z_{1})\neq f_{2}(z_{1}).$ That is, for any of the proportions
$\frac{L_{x}(t)}{L(t)}$ for $x\in[0,\omega),$ the remaining years
to live is not equal to $y_{1}.$ 

We will continue matching $\frac{P_{y_{2}}(t)}{P(t)}$ for some $y_{2}\neq y_{1}$
and $y_{2}\in[0,\omega)$ with $\frac{L_{x}(t)}{L(t)}$ for each $x\in A$
except for $x=y_{1}.$ If there is a value of $\frac{L_{x}(t)}{L(t)}$
that equals $\frac{P_{y_{2}}(t)}{P(t)},$ we call the corresponding
proportion in life table population as $\frac{L_{y_{2}}(t)}{L(t)}$
(say, $f_{2}(y)$). $\frac{L_{y_{2}}(t)}{L(t)}$ is the proportion
of life table sub-population who have $y_{2}$ years to live. That
is,

\begin{equation}
g_{1}(y_{2})=\frac{P_{y_{2}}(t)}{P(t)}=\frac{L_{y_{2}}(t)}{L(t)}=f_{2}(y_{2}).\label{eq:4}
\end{equation}

Suppose $\frac{P_{y_{2}}(t)}{P(t)}$ does not equal to any of the
proportions $\frac{L_{x}(t)}{L(t)}$ for $x\in A$, then we denote
$y_{2}$ in (\ref{eq:4}) by $z_{2}$ (if $z_{1}$ already arises
in an earlier situation such that $g_{1}(z_{1})\neq f_{2}(z_{1}))$,
then we write $g_{1}(z_{2})\neq f_{2}(z_{2}).$ However, if $g_{1}(y_{1})=f_{2}(y_{1})$
exists but $\frac{P_{y_{2}}(t)}{P(t)}\neq\frac{L_{x}(t)}{L(t)}$ $\forall x\in A$,
then $y_{2}$ in (\ref{eq:4}) we denote as $z_{1}$ and write $g_{1}(z_{1})\neq f_{2}(z_{1}).$

\begin{figure}
\includegraphics[scale=0.6]{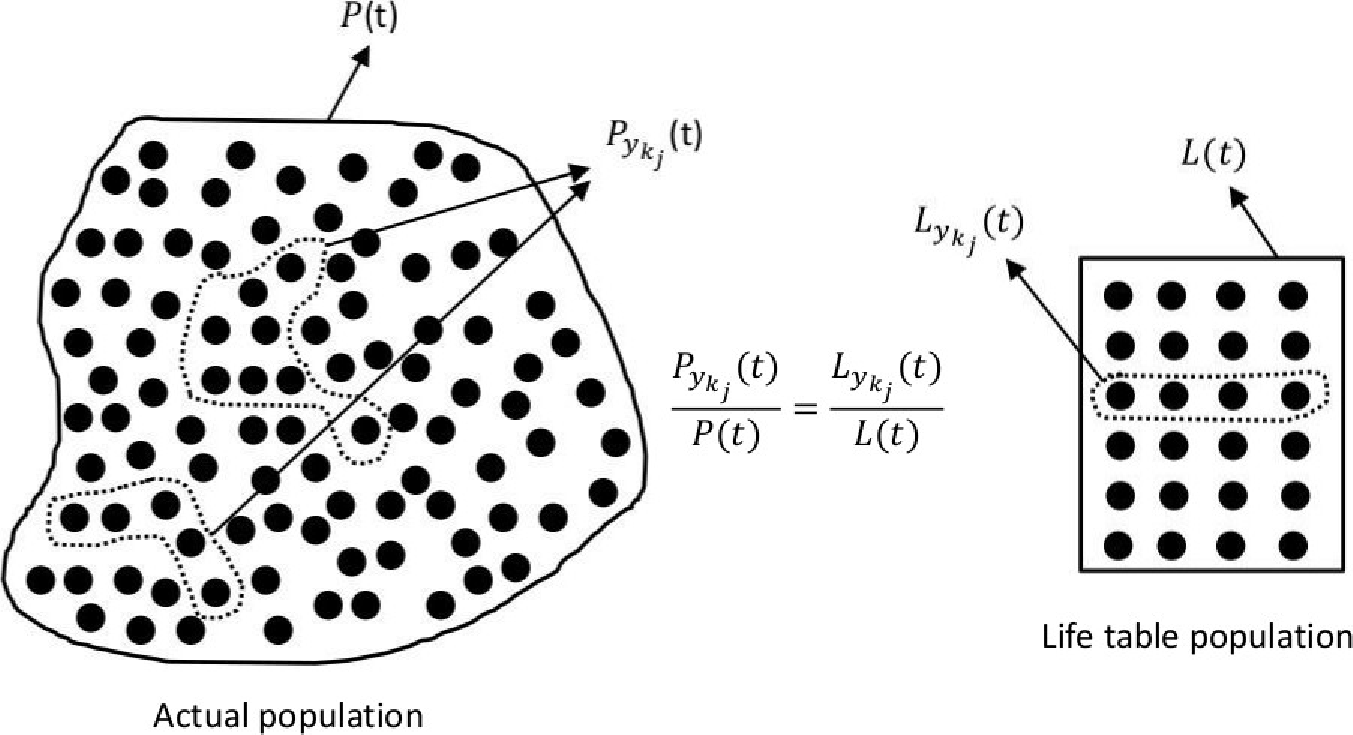}

\caption{\label{fig:Fractions-of-population}proportions of population in actual
and in life table. $P_{y_{k_{j}}}(t)$ be its sub-population who are
at age $y_{k_{j}}$, where $P(t)$ is the total actual population
and $L_{y_{k_{j}}}(t)$ is the life table population at time $t$
whose remaining years to live is $y_{k_{j}}$ and $L(t)$ is total
life table population at $t$.}
\end{figure}

\begin{figure}
\includegraphics[scale=0.7]{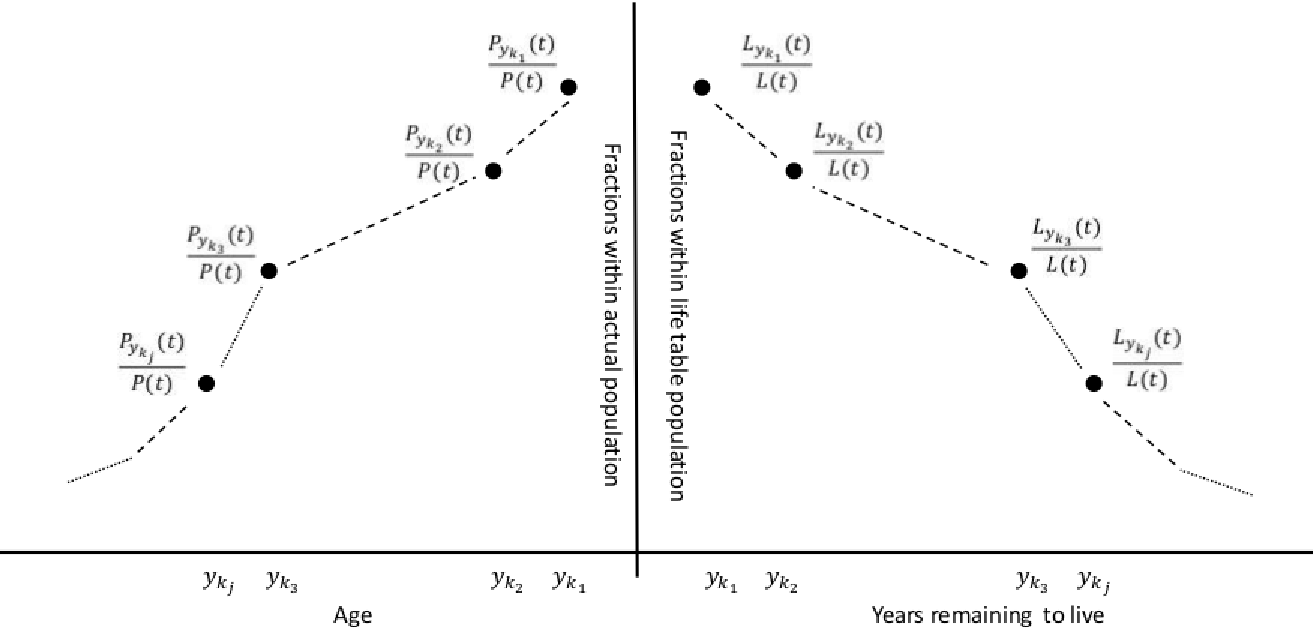}

\caption{\label{fig:Stationary-component-Figure}Stationary component of a
randomly selected large population. proportional sub-populations at
ages $y_{k_{1}},y_{k_{2}},...$ out of actual total population are
identical with proportional life table sub-populations who have remaining
years $y_{k_{1}},y_{k_{2}},...$.}
\end{figure}

\begin{figure}
\includegraphics[scale=0.7]{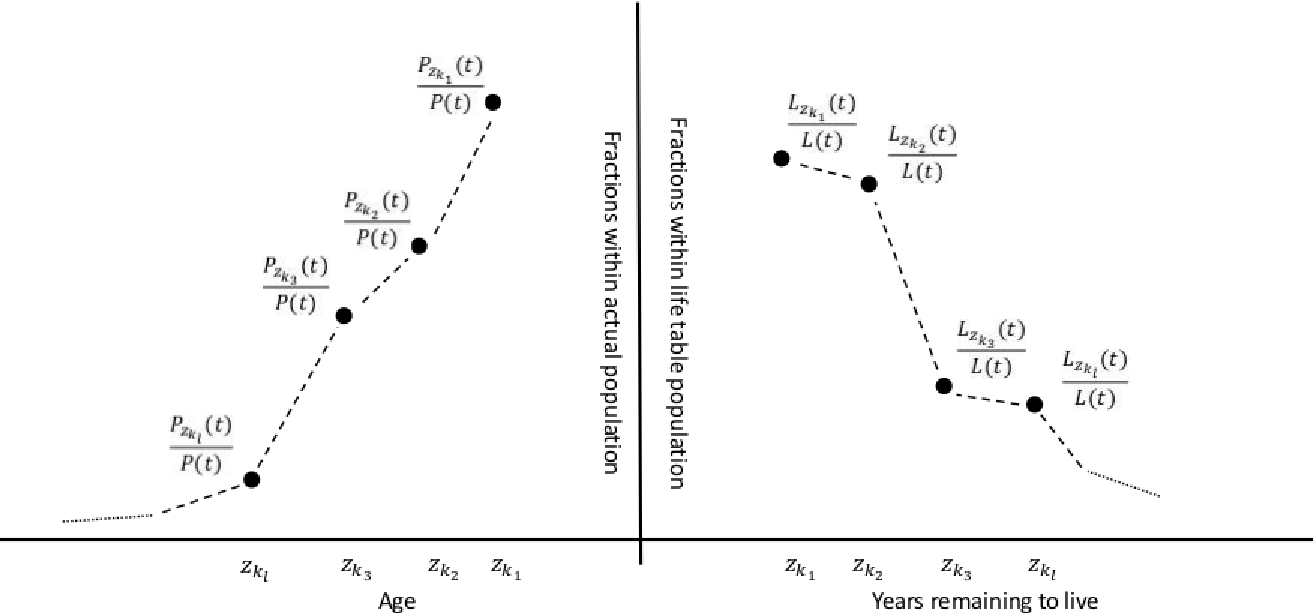}

\caption{\label{fig:Non-stationary-component}Non-stationary component of a
randomly selected large population. proportional sub-populations at
ages $z_{k_{1}},z_{k_{2}},...$ out of actual total population are
not identical with any of the life table proportional sub-populations
whose remaining years are $z_{k_{1}},z_{k_{2}},...$ }
\end{figure}

Similarly, for a randomly selected age $y_{i}\in A$, the proportion
$\frac{P_{y_{i}}(t)}{P(t)}$ (i.e. $g_{1}(y_{i})$) is matched with
the proportion $\frac{L_{x}(t)}{L(t)}$ $\forall x\in A.$ If there
is a value of $\frac{P_{y_{i}}(t)}{P(t)}$ that matches with $\frac{L_{x}(t)}{L(t)}$
then we denote it by $f_{2}(y_{i}),$ otherwise we denote it by $f_{2}(z_{i})$
(if previous $z$ value in the order of unmatched proportions was
denoted as $z_{i-1}$ for $i=2,3,...$). Through this procedure we
will match for all the values of $y_{i}\in A,$ and decide whether
or not $g_{1}(y_{i})$ is equal to the $f_{2}(y_{i}).$ The criteria
is, for an age $y_{i}$ in a randomly selected large population (actual
population), if the value of the proportion $\frac{P_{y_{i}}(t)}{P(t)}$
is equal to any of the life table sub-population proportions $\frac{L_{x}(t)}{L(t)}$
$\forall x\in A$ whose remaining years to live is exactly $y_{i},$
then, $g_{1}(y_{i})=f_{2}(y_{i}).$ 

{\large{}Existence of two sets $\{y_{k_{j}}\}$ and $\{z_{k_{j}}\}:$}{\large\par}

In general, at each age $x\in A$, we can check whether

\begin{equation}
\frac{P_{x}(t)}{P(t)}=\frac{L_{x}(t)}{L(t)}\text{ holds}\label{eq:twi-five}
\end{equation}

or

\begin{equation}
\frac{P_{x}(t)}{P(t)}\neq\frac{L_{x}(t)}{L(t)}\text{ holds.}\label{eq:two-six}
\end{equation}

Suppose we start this procedure from age $0.$ At $x=0$, either (\ref{eq:twi-five})
holds or (\ref{eq:two-six}) holds but not both. If (\ref{eq:twi-five})
holds then, let us denote $y_{k_{1}}$ for $0$, or if (\ref{eq:two-six})
holds then, let us denote $z_{k_{1}}$ for $0$. Hence, at age $0$
one of the $y_{k_{1}}$ or $z_{k_{1}}$ exists, and 

\begin{equation}
\left\{ y_{k_{1}}\right\} \cup\left\{ z_{k_{1}}\right\} =\left\{ 0\right\} \label{eq:singleton-0}
\end{equation}

Now, let us consider age $a_{1}>0$ for $a_{1}\in A$. At age $x=a_{1}$,
either (\ref{eq:twi-five}) holds or (\ref{eq:two-six}) holds but
not both. If (\ref{eq:twi-five}) holds at $x=a_{1}$, and $y_{k_{1}}$
exists, then let us denote $y_{k_{2}}$ for $a_{1}$. If (\ref{eq:twi-five})
holds at $x=a_{1}$ and $z_{k_{1}}$ exists, then let us denote $y_{k_{1}}$
for $a_{1}$. If (\ref{eq:two-six}) holds at $x=a_{1}$ and $y_{k_{1}}$
exist, then let us denote $z_{k_{1}}$ for $a_{1}$. If (\ref{eq:two-six})
is true at $x=a_{1}$ and $z_{k_{1}}$ exists, then let us denote
$z_{k_{2}}$ for $a_{1}$. From the arguments constructed so far,
we have shown the existence of one of the following sets:

\begin{equation}
\left\{ \{y_{k_{1}},y_{k_{2}}\},\{z_{k_{1}},y_{k_{1}}\},\{y_{k_{1}},z_{k_{1}}\},\{z_{k_{1}},z_{k_{2}}\}\right\} \label{eq:existence2}
\end{equation}

The union of all the elements of the set (\ref{eq:existence2}) is

\begin{equation}
\left\{ \{y_{k_{1}},y_{k_{2}}\}\cup\{z_{k_{1}},y_{k_{1}}\}\cup\{y_{k_{1}},z_{k_{1}}\}\cup\{z_{k_{1}},z_{k_{2}}\}\right\} =\left\{ 0,a_{1}\right\} .\label{eq:union=00007B0-a1=00007D}
\end{equation}

Now, let us consider age $a_{2}$ for $a_{2}>a_{1}>0$ and $a_{2}\in A$.
At age $x=a_{2}$, either (\ref{eq:twi-five}) holds or (\ref{eq:two-six})
holds but not both. 

$ $

Suppose (\ref{eq:twi-five}) holds at $x=a_{2}$. For $x=0$ and $x=a_{1}$,
there is a possibility of occurrence of one of the four sets of (\ref{eq:existence2})
. All possible combinatorics of $x$ values are explained below:

$(i)$ (\ref{eq:twi-five}) holds at $x=a_{2}$ and $\left\{ y_{k_{1}},y_{k_{2}}\right\} $
exists, then let us denote $y_{k_{3}}$ for $a_{2}$. 

$(ii)$ (\ref{eq:twi-five}) holds at $x=a_{2}$ and $\{z_{k_{1}},y_{k_{1}}\}$
exists, then let us denote $y_{k_{2}}$ for $a_{2}$. 

$(iii)$ (\ref{eq:twi-five}) holds at $x=a_{2}$ and $\{y_{k_{1}},z_{k_{1}}\}$
exists, then let us denote $y_{k_{2}}$ for $a_{2}$. 

$(iv)$ (\ref{eq:twi-five}) holds at $x=a_{2}$ and $\{z_{k_{1}},z_{k_{2}}\}$
exists, then let us denote $y_{k_{1}}$ for $a_{2}$. 

$ $

Suppose (\ref{eq:two-six}) holds at $x=a_{2}$. For $x=0$ and $x=a_{1}$,
there is a possibility of occurrence of one of the four sets of (\ref{eq:existence2}).
All possible combinatorics of $x$ values are explained below:

$(v)$ (\ref{eq:two-six}) holds at $x=a_{2}$ and $\left\{ y_{k_{1}},y_{k_{2}}\right\} $
exists, then let us denote $z_{k_{1}}$ for $a_{2}$. 

$(vi)$ (\ref{eq:two-six}) holds at $x=a_{2}$ and $\{z_{k_{1}},y_{k_{1}}\}$
exists, then let us denote $z_{k_{2}}$ for $a_{2}$. 

$(vii)$ (\ref{eq:two-six}) holds at $x=a_{2}$ and $\{y_{k_{1}},z_{k_{1}}\}$
exists, then let us denote $z_{k_{2}}$ for $a_{2}$. 

$(viii)$ (\ref{eq:two-six}) holds at $x=a_{2}$ and $\{z_{k_{1}},z_{k_{2}}\}$
exists, then let us denote $z_{k_{3}}$ for $a_{2}$. 

$ $

Through $(i)$ to $(viii)$ we have shown the existence of one of
the following sets: 

\begin{equation}
\text{\ensuremath{\left\{  \begin{array}{c}
 \{y_{k_{1}},y_{k_{2}},y_{k_{3}}\},\{z_{k_{1}},y_{k_{1}},y_{k_{2}}\},\{y_{k_{1}},z_{k_{1}},y_{k_{2}}\},\{z_{k_{1}},z_{k_{2}},y_{k_{1}}\}\\
 \{y_{k_{1}},y_{k_{2}},z_{k_{1}}\},\{z_{k_{1}},y_{k_{1}},z_{k_{2}}\},\{y_{k_{1}},z_{k_{1}},z_{k_{2}}\},\{z_{k_{1}},z_{k_{2}},z_{k_{3}}\} 
\end{array}\right\} } }\label{eq:existence-a2}
\end{equation}

The union of all the elements (i.e. $2^{3}$ number of sets) of the
set (\ref{eq:existence-a2}) is 
\begin{equation}
\left\{ 0,a_{1},a_{2}\right\} .\label{eq:union=00007B0,a1,a2=00007D}
\end{equation}

$ $

The number of possible sets at age $x=a_{3}$ once we construct similar
to the previous arguments would be double the number of sets of (\ref{eq:existence-a2}),
which is $2^{4}$ number of sets. These $2^{4}$ are given below:

\begin{equation}
\text{\ensuremath{\left\{  \begin{array}{c}
 \{y_{k_{1}},y_{k_{2}},y_{k_{3}},y_{k_{4}}\},\{z_{k_{1}},y_{k_{1}},y_{k_{2}},y_{k_{3}}\},\{y_{k_{1}},z_{k_{1}},y_{k_{2}},y_{k_{3}}\},\{z_{k_{1}},z_{k_{2}},y_{k_{1}},y_{k_{2}}\}\\
 \{y_{k_{1}},y_{k_{2}},z_{k_{1}},y_{k_{3}}\},\{z_{k_{1}},y_{k_{1}},z_{k_{2}},y_{k_{2}}\},\{y_{k_{1}},z_{k_{1}},z_{k_{2}},y_{k_{2}}\},\{z_{k_{1}},z_{k_{2}},z_{k_{3}},y_{k_{1}}\}\\
 \{y_{k_{1}},y_{k_{2}},y_{k_{3}},z_{k_{1}}\},\{z_{k_{1}},y_{k_{1}},y_{k_{2}},z_{k_{2}}\},\{y_{k_{1}},z_{k_{1}},y_{k_{2}},z_{k_{2}}\},\{z_{k_{1}},z_{k_{2}},y_{k_{1}},z_{k_{3}}\}\\
 \{y_{k_{1}},y_{k_{2}},z_{k_{1}},z_{k_{2}}\},\{z_{k_{1}},y_{k_{1}},z_{k_{2}},z_{k_{3}}\},\{y_{k_{1}},z_{k_{1}},z_{k_{2}},z_{k_{3}}\},\{z_{k_{1}},z_{k_{2}},z_{k_{3}},z_{k_{4}}\} 
\end{array}\right\} } }\label{eq:existence of a3}
\end{equation}

The union of $2^{4}$ sets in (\ref{eq:existence of a3}) is

\[
\left\{ 0,a_{1},a_{2},a_{3}\right\} .
\]

Note that the elements of (\ref{eq:existence of a3}) are drawn from
the set 

\[
\{y_{k_{1}},y_{k_{2}},y_{k_{3}},y_{k_{4}},z_{k_{1}},z_{k_{2}},z_{k_{3}},z_{k_{4}}\}
\]

which is of size $2.4=8.$

In a similar way, by the induction, the number of possible sets at
age $x=a_{j}$ for $a_{j}>a_{j-1}>...>a_{1}>a_{0}$ would be $2^{j+1}$.
Each of the set in the collection of $2^{j+1}$ will be of size $j+1$.
These $j+1$ elements are drawn from the unique combinations of $2j+2$
elements, viz, 

\begin{equation}
\left\{ y_{k_{1}},y_{k_{2}},...,y_{k_{j+1}},z_{k_{1}},z_{k_{2}},...,z_{k_{j+1}}\right\} \label{eq:existence-aj}
\end{equation}
The union of the $2^{j+1}$ sets would be

\begin{equation}
\left\{ 0,a_{1},a_{2},...,a_{j}\right\} .\label{eq:a1...aj}
\end{equation}

When $a_{j}=\omega$ in (\ref{eq:a1...aj}), it becomes the set $A.$
The elements of $A$ were drawn from 

\[
\left\{ y_{k_{1}},y_{k_{2}},...,\omega,z_{k_{1}},z_{k_{2}},...,\omega\right\} 
\]

Here,

\begin{equation}
\bigcup_{p=1}^{j}\left\{ y_{k_{p}}\right\} \cup\left\{ z_{k_{p}}\right\} =A,\label{eq:5}
\end{equation}

and corresponding sub-populations' totals for the individuals who
are all at ages $\left\{ y_{k_{j}}\right\} $ and $\left\{ z_{k_{l}}\right\} $,
are $P_{y_{k_{j}}}(t)$ and $P_{z_{k_{l}}(t)},$ respectively. Due
to (\ref{eq:5}), we can write,

\[
\Sigma_{j=1}^{\omega}P_{y_{k_{j}}}(t)+\Sigma_{l=1}^{\omega}P_{z_{k_{l}}(t)}=P(t),
\]

where $\Sigma_{j=1}^{\infty}P_{y_{k_{j}}}(t)$ is $M(t)$ formed by
satisfying the SPI at the ages $\left\{ y_{k_{j}}\right\} $, and
$\Sigma_{l=1}^{\infty}P_{z_{k_{l}}(t)}$ is $N(t)$ formed by not
satisfying the SPI at the ages $\left\{ z_{k_{l}}\right\} $. Hence,
the proof. 

\begin{figure}

\includegraphics[scale=0.65]{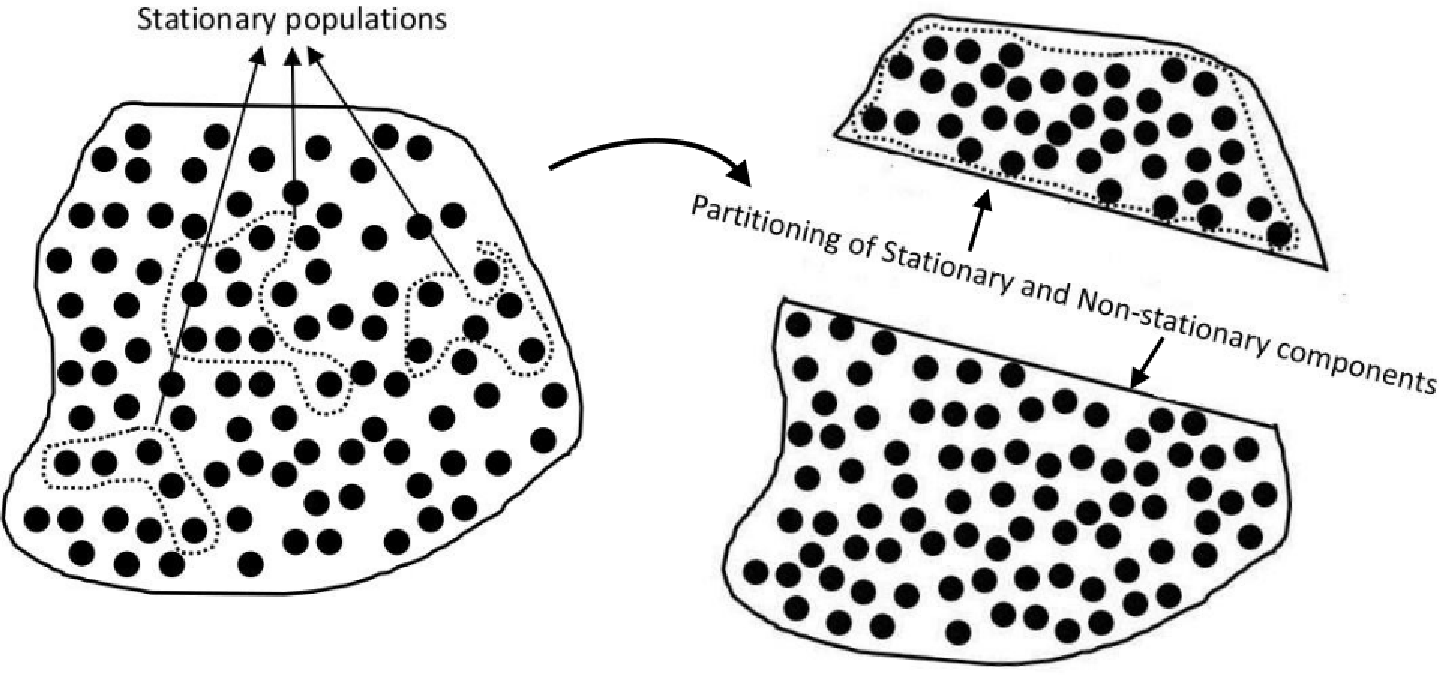}\caption{\label{fig:Partitionsfigure}Partitions into stationary and non-stationary
components of a randomly selected large population. All the populations
within the dotted lines on the left hand side form the stationary
component and others make non-stationary component of the total population
as shown in the right side of the Figure. Note that partitioning is
not about geographical partition. }

\end{figure}
\end{proof}
\begin{rem}
Since we are calculating the proportion of sub-populations at each
age of the actual population, we might come across values of these
proportions calculated at two or more ages that could be identical.
All the ages with identical proportions would fall within the same
component of the population, i.e. stationary or non-stationary by
the construction explained in the proof of the Theorem \ref{prop:Stationary-Population-Identity}.
\end{rem}

$ $
\begin{rem}
The partition theorem can be extended to more than two partitions
if there are multiple decrement life tables available for a randomly
selected large population.
\end{rem}

\section{Conclusions}

The main theorem stated and proved is the first such observation in
the literature. Moreover, we have not come across in the literature
where life table identity was being used to relate actual populations
and applied to decide stationary and non-stationary components of
a large randomly selected actual population. The partitioning procedure
described in this work can be used to decide what proportion of the
population is stationary and what proportion is not by considering
the world population as a whole as one unit, individual countries,
continents and groups of countries, etc, We can also apply this procedure
to test the stationary and non-stationary status of all sub-regions
of a large country. 

The partition theorem stated and proved is not an improvement of any
previously shown results in stationary populations. The statement
of partition theorem in the article is original and the method demonstrated
whether a component of a large population is stationary or non-stationary
has not existed. A given large population or its sub-populations showing
oscillatory behavior of transitioning from stationary to non-stationary
and vice versa were earlier shown in \cite{RaoCareyOSPI}. 

With many countries in the world approaching or at replacement levels
but often with widely varying if not unique age distributions, the
partitioning theory outlined here has the potential to provide new
metrics on age structure in particular and on overall population dynamics
in general. Metrics to measure population stability status between
two populations were proposed in \cite{RaoASRSNotices}. Both of these
can be used for both between-country comparisons as well as for projections
into the future at both regional, national and global levels.

Acknowledgments: J.R. Carey (U.C. Davis) provided appreciation and
encouragement when the author posed the statement of the partition
theorem for populations for the first time in 2018. This inspired
the author to finish the proof. S. Tuljapurkar (Stanford University)
provided helpful comments on the previous draft. Two reviewers provided
very helpful and constructive comments that very much helped in revising
the article. I am greatly thankful to all. ASRS Rao has no funding
support to disclose that is related to this project.

\end{document}